\documentclass[11pt]{article}

\usepackage[a4paper,margin=1in]{geometry}
\usepackage{microtype}

\usepackage{amsmath,amsfonts,amssymb}

\usepackage{graphicx}
\usepackage{float}
\usepackage{booktabs,tabularx}

\usepackage{enumitem}
\usepackage{authblk}
\usepackage[hidelinks]{hyperref}

\title{Intensity-based Segmentation of Tissue Images Using a U-Net with a Pretrained ResNet-34 Encoder: Application to Mueller Microscopy}

\author[1$^{*\dagger}$]{Sooyong Chae}
\author[2$^{*}$]{Dani Giammattei}
\author[2]{Ajmal Ajmal}
\author[2]{Junzhu Pei}
\author[2]{Amanda Sanchez}
\author[2]{Tananant Boonya-ananta}
\author[2]{Andres Rodriguez}
\author[1,2]{Tatiana Novikova}
\author[2,3]{Jessica Ramella-Roman}

\affil[1]{LPICM, CNRS, \'Ecole Polytechnique, IP Paris, 91120 Palaiseau, France}
\affil[2]{Department of Biomedical Engineering, Florida International University, Miami, FL 33174, USA}
\affil[3]{Department of Ophthalmology, Herbert Wertheim College of Medicine, Florida International University, Miami, FL 33199, USA}
\affil[ ]{$^{*}$These authors contributed equally to this work.}
\affil[ ]{$^{\dagger}$\texttt{sooyong.chae@polytechnique.edu}}
\date{}

\begin{document}

\maketitle

\begin{abstract}
Manual annotation of the images of
thin tissue sections remains time-consuming step in Mueller 
microscopy and limits its scalability. We present a novel automated approach using only the total intensity M$_{11}$ element of the Mueller matrix as an input to a U-Net architecture with a pretrained ResNet-34 encoder. The network was trained to distinguish four classes in the images of murine uterine cervix sections: background, internal os, cervical 
tissue, and vaginal wall. With only 70 cervical tissue sections
, the model achieved 89.71\% pixel accuracy and 80.96\% mean tissue Dice coefficient on the held-out test dataset. Transfer learning from ImageNet enables accurate segmentation despite limited size of training dataset typical of specialized biomedical imaging. This intensity-based framework requires minimal preprocessing and is readily extensible to other imaging modalities and tissue types, with publicly available graphical annotation tools for practical deployment.
\end{abstract}


\section{Introduction}
\label{sec:m11_filter_introduction}
Imaging Mueller polarimetry has been extensively explored for biomedical applications 
~\cite{ghosh2011tissue, 
Booth,
qi2017mueller, ramella2022polarized}. 
For quantitative analysis in Mueller matrix (MM) microscopy, accurate anatomical segmentation is a necessary for obtaining structure-specific polarimetric maps.
Physical realizability (PR) test of the MM 
~\cite{cloude1986group, Pogudin2024} remains the first step of experimental polarimetric data post-processing. As demonstrated in our previous work~\cite{chae2025machine}, the PR test 
serves as an effective physics-informed filter for the MM images of bulk tissues, naturally retaining tissue regions, while excluding background and non-physical artifacts.

\begin{figure}[htbp]
\centering
\includegraphics[width=0.8\linewidth]{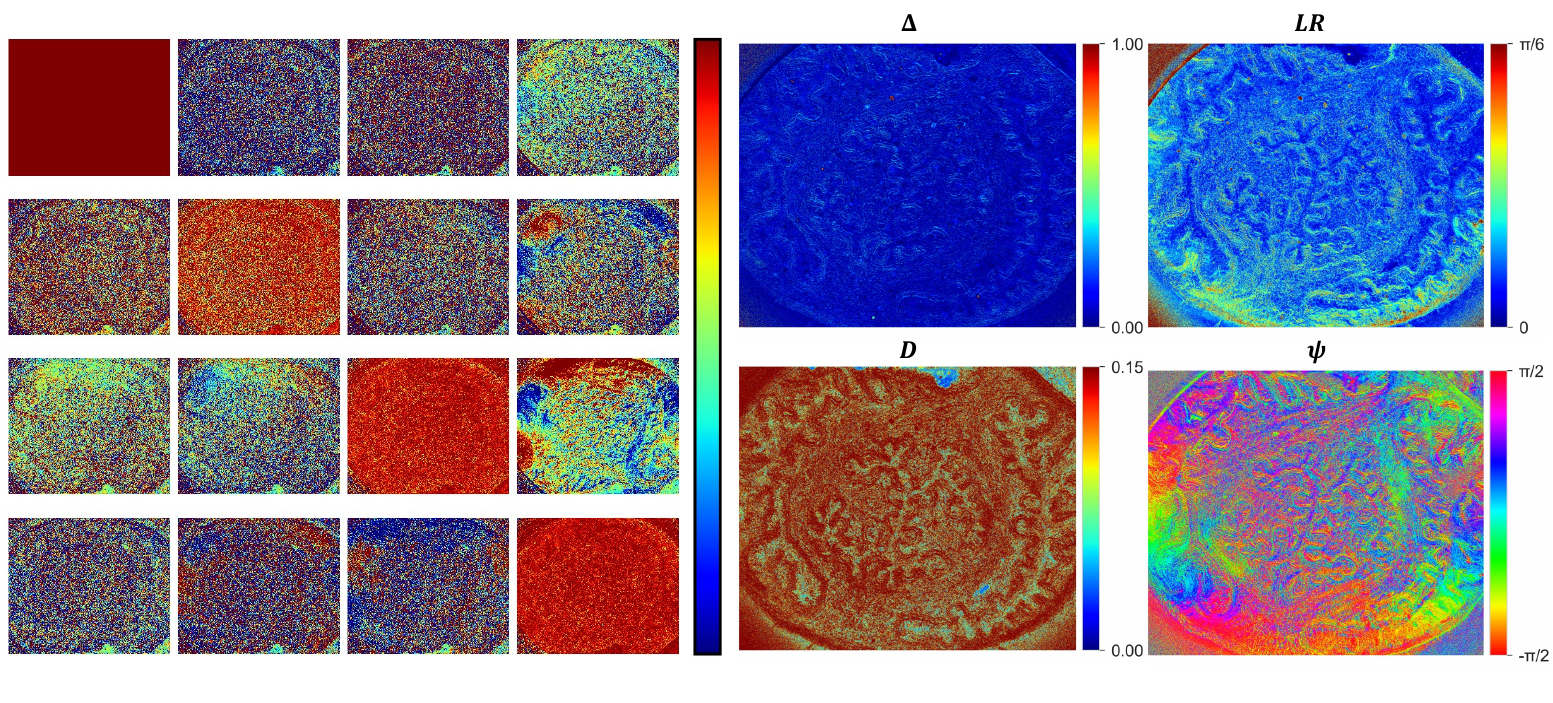}
\caption{Images of murine uterine cervix thin section (Day 18 of the gestation). \textbf{Left:} 
MM images. 
Color bar: $[0,1]$ (diagonal elements), $[-0.1,0.1]$(off-diagonal elements). \textbf{Right:} Derived maps of polarimetric parameters using Lu-Chipman decomposition~\cite{LC
}: depolarization $\Delta$ (dimensionless), diattenuation $D$ (dimensionless), linear retardance $LR$ (rad), and optical axis azimuth $\psi$ (rad).}
\label{fig:m11_filter_thin_mm_sample}
\end{figure}

However, thin tissue sections exhibit weak polarimetric response 
with diagonal MM elements close to unity and reduced image contrast, making PR-based filtering less effective. Despite this reduced contrast, polarimetric information remains detectable in the polarimetric images of thin sections: linear retardance $LR$ and optical axis orientation $\psi$ exhibit spatial patterns corresponding to anisotropic tissue organization (Fig.~\ref{fig:m11_filter_thin_mm_sample}). For example, several studies have demonstrated that polarimetric imaging characterizes cervical collagen organization~\cite{chue2017use, chue2018use} and detects its remodeling during pregnancy~\cite{lee2021mueller, ramella2024quantitative}.

To enable structure-specific polarimetric analysis, accurate segmentation of anatomical regions is required~\cite{Lee2018, Lee2022}. While the M$_{11}$ element provides sufficient morphological contrast for this task, manual annotation of tissue regions and anatomical structures remains tedious and time-consuming, limiting scalability.
Deep learning offers automated segmentation, but typically requires hundreds to thousands of annotated samples~\cite{ronneberger2015u} — a limitation when sample acquisition involves animal protocols, histological preparation, and custom instrumentation.
We demonstrate that, 
when imaging cervical samples, 
the transfer learning from ImageNet-pretrained networks enables accurate tissue segmentation from MM data using only 70 annotated sample images. 

We employ a U-Net architecture~\cite{ronneberger2015u} with pretrained ResNet-34 encoder~\cite{he2016deep} for pixel-wise segmentation using only the image of M$_{11}$ element of MM. This intensity-based approach requires minimal preprocessing compared to full MM decomposition and leverages natural image features learned from ImageNet to identify structural patterns in sample images. We provide an intuitive graphical interface for annotation and model deployment, enabling practical implementation without extensive programming.
\vskip 0.05in

\section{Instrumentation}
\label{sec:m11_filter_instrumentation}
MM images of thin cervical tissue sections were acquired using a custom transmission-mode MM polarimeter built on a commercial trinocular microscope (ME580T-PZ-2L, AmScope). Samples were illuminated with a 9W stabilized broadband source (SLS201L, Thorlabs) filtered at 550 nm (FB550-10-1, Thorlabs). Complete $4\times4$ MM images were reconstructed from 16 intensity images acquired under different polarization states generated by a motorized Polarization State Generator (PSG) and analyzed by a Polarization State Analyzer (PSA). Each comprised a linear polarizer (LPVISC100, Thorlabs) and motorized quarter-wave plate (AQWP10M-580, Thorlabs) on a rotation stage (PRM1Z8, Thorlabs). Samples were imaged with a 5$\times$ objective (NA 0.13, 4 mm field of view) and recorded with a 16-bit sCMOS camera (PCO.edge 5.5) with an image size of $2560 \times 2160$ pixels and resolution of 
$\sim 1.5\text{ }\mu m$ per pixel.
\vskip 0.05in

\section{Sample Collection and Annotation}
\label{subsec:m11_filter_acquisition}
\begin{figure}[h!]
\centering
\includegraphics[width=0.7\linewidth]{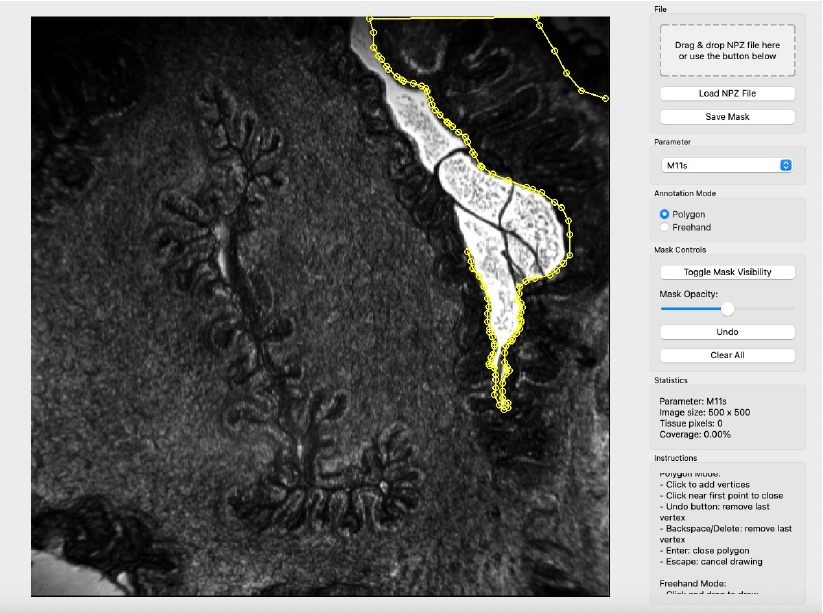}
\caption{Screenshot of the custom Tissue Annotation Tool GUI. The interface features an interactive canvas for visualizing the normalized $M_{11}$ intensity map. The control panel enables users to switch between vertex-based Polygon and Freehand drawing modes, adjust mask layer opacity, and define the semantic classes (e.g., internal os, vaginal wall)}
\label{fig:m11_filter_gui}
\end{figure}

%
%
Cervical tissue was selected as a representative anatomical challenge due to its heterogeneous structure and the morphological changes it undergoes across gestation. The dataset comprised 70 thin ($\sim$50 µm) sections of murine uterine cervix across early to late gestation periods (Days 0 -- 18).
To generate ground truth segmentation masks from this small dataset, we developed a custom annotation tool using PyQt5~\cite{pyqt5}. The software displays pixel-wise normalized M$_{11}$ intensity images and provides interactive polygon and freehand drawing modes for delineating the regions of interest (Fig.~\ref{fig:m11_filter_gui}). 

\begin{figure}[h]
\centering
\includegraphics[width=0.7\linewidth]{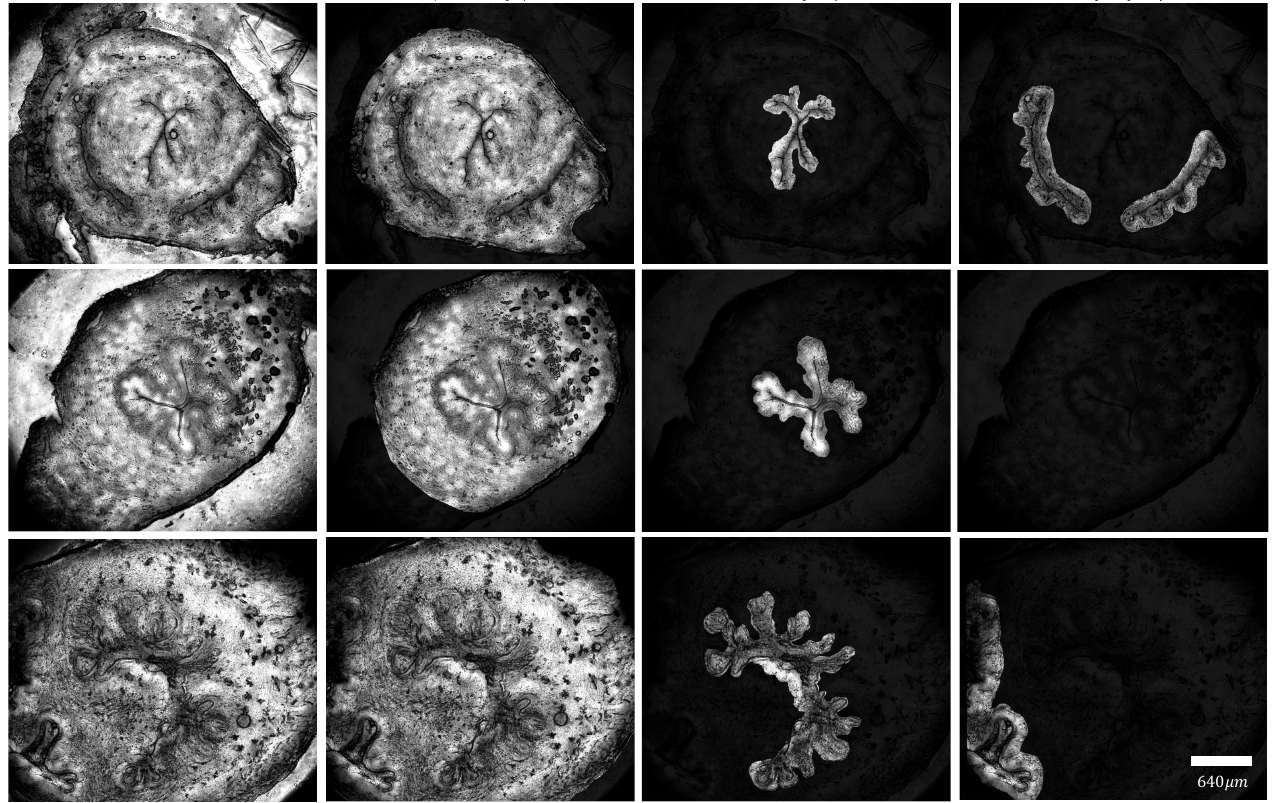}
\caption{Representative 
images 
of cervical tissue sections.
\textbf{Left to Right}: M$_{11}$ intensity image, tissue mask, OS mask, and vaginal wall mask. \textbf{Top}
: Clear OS and vaginal wall structures. \textbf{Middle}
: 
Absent vaginal walls.
\textbf{Bottom}
: Irregular internal vaginal wall shapes and ambiguous tissue borders.
}
\label{fig:m11_filter}
\end{figure}

Separate binary masks were created for general tissue, internal os, and vaginal wall, then combined into four-class maps with priority hierarchy: vaginal wall (class 3) $>$ OS (class 2) $>$ tissue (class 1) $>$ background (class 0). The tool is distributed as standalone executables for macOS and Windows to ensure accessibility without programming expertise (see Code Availability).
Manual annotation shows notable morphological variability across samples (Fig.~\ref{fig:m11_filter}). 

Tissue boundaries exhibited ambiguity from preparation-related damage, internal os shapes varied substantially, and vaginal wall presence ranged from complete to absent depending on sectioning depth. This heterogeneity, combined with the small sample size, presents a challenging test case for deep learning generalization.

The dataset was stratified by acquisition day and split into training (49 samples, 70\%), validation (10 samples, 15\%), and test (11 samples, 15\%) sets to ensure generalization and prevent data leakage between experimental sessions.

\section{Methods}
\begin{figure}[htbp]
\centering
\includegraphics[width=0.7\linewidth]{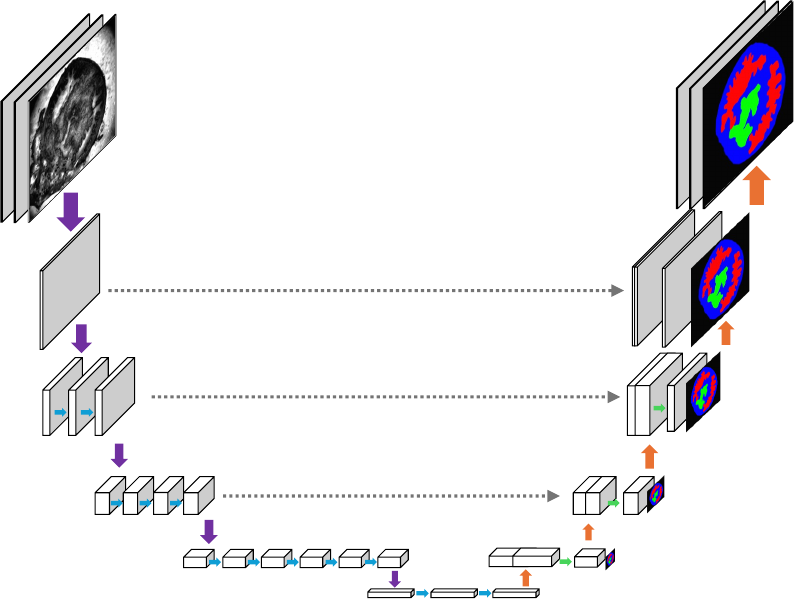}
\caption{U-Net with a ResNet-34 encoder for M$_{11}$ 
tissue image segmentation. The encoder downsamples the input (purple arrows) with internal cross-layer connections (blue arrows). Skip connections (gray dashed arrows) pass high-resolution features to the decoder, which upsamples the features (orange arrows) with internal cross-layer connections (green arrows). 
Spatial resolution decreases from $256\times256$ to $16\times16$ at the bottleneck and increases to $512\times512$ at the output.}
\label{fig:m11_filter_architecture}
\end{figure}

%
We employed a U-Net architecture~\cite{ronneberger2015u} with a pretrained ResNet-34 encoder~\cite{he2016deep} for pixel-wise classification into four classes: background, general tissue, internal os, and vaginal wall.
The model architecture is shown in Fig~\ref{fig:m11_filter_architecture}.
The encoder (initialized with ImageNet-1K v1 pretrained weights~\cite{deng2009imagenet}) progressively downsamples the input ($512\times512$ pixels) through five blocks with channel dimensions [64, 64, 128, 256, 512]. Input M$_{11}$ images were resized to $512\times512$ pixels using bilinear interpolation, while corresponding masks were resized using nearest-neighbor interpolation to preserve discrete class labels~\cite{long2015fully}. 

Single-channel M$_{11}$ images were replicated to three channels and normalized using ImageNet statistics (mean=[0.485, 0.456, 0.406], std=[0.229, 0.224, 0.225]) to ensure compatibility with pretrained weights. The decoder reconstructs spatial resolution via skip connections and bilinear upsampling, applying convolution-BatchNorm-ReLU blocks~\cite{ioffe2015batch} to produce four-channel output logits.

Training used a combined loss ($\mathcal{L}_{\text{total}} = 0.5 \cdot \mathcal{L}_{\text{CE}} + 0.5 \cdot \mathcal{L}_{\text{Dice}}$) balancing cross-entropy and Dice loss~\cite{milletari2016v}, with AdamW optimizer~\cite{loshchilov2017decoupled} (learning rate $1\times10^{-4}$, weight decay $1\times10^{-5}$), batch size of 8, and 50 epochs. The best model was selected based on minimum validation loss. Data augmentation included random horizontal/vertical flips, 90-degree rotations, and intensity jittering ($\pm15\%$). The dataset was stratified by acquisition day (70\% training, 15\% validation, 15\% testing) for generalization. All experiments were conducted using PyTorch~\cite{paszke2019pytorch} with mixed-precision training~\cite{micikevicius2017mixed} on NVIDIA GPU (RTX5000 ADA, 16Go DDR6). 
\vskip 0.05in
\section{Results}

\begin{figure}[H]
\centering
\includegraphics[width=\linewidth]{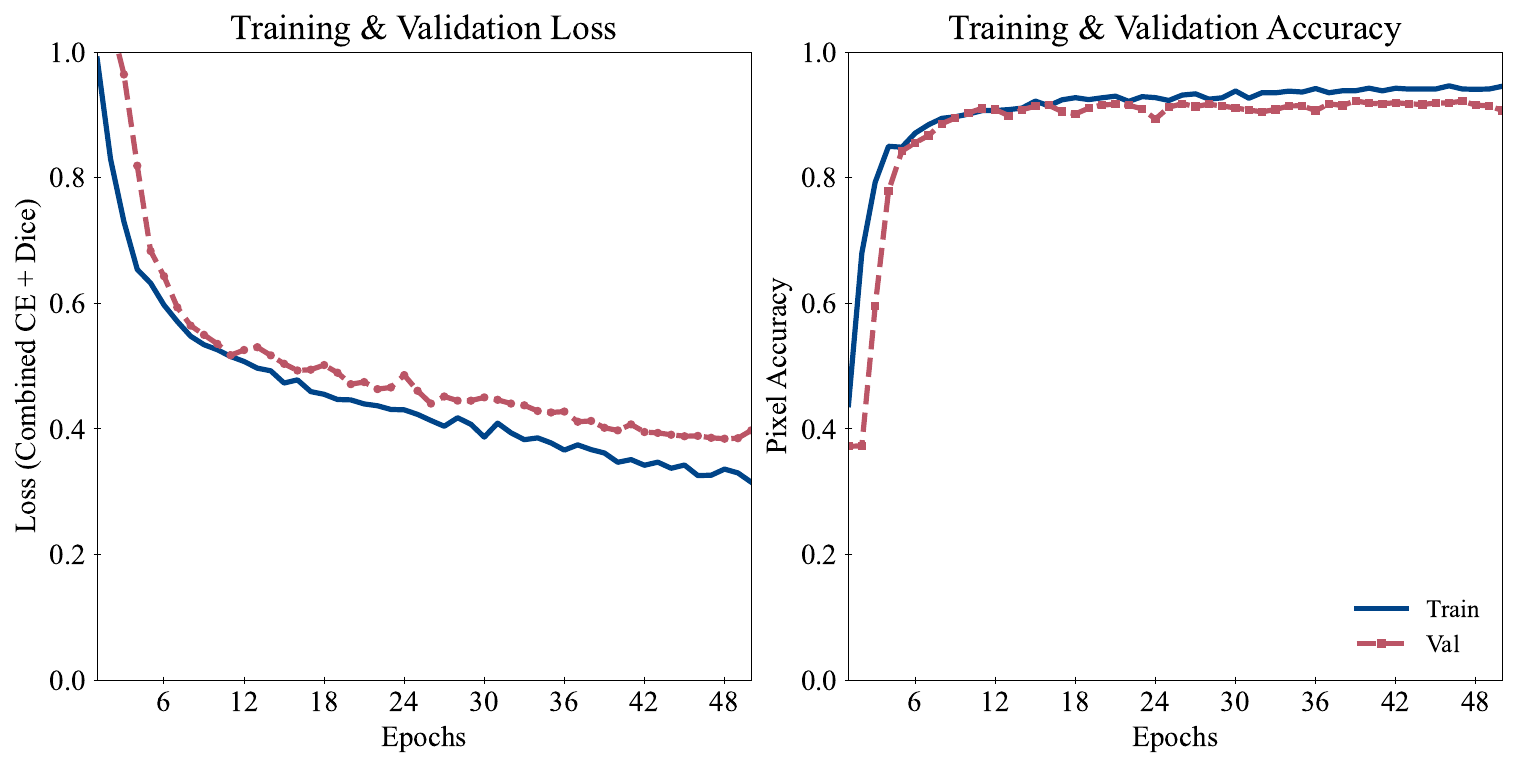}
\caption{Training metrics over 50 epochs. \textbf{Left:} combined loss showing steady convergence. \textbf{Right}: pixel accuracy demonstrating robust generalization.}
\label{fig:m11_training_curves}
\end{figure}

The model achieved 91.72\% training accuracy and 89.66\% validation accuracy after 50 epochs, with validation loss converging to 0.374 (Fig.~\ref{fig:m11_training_curves}). Close alignment between training and validation curves throughout training indicates effective generalization without overfitting.

On the held-out test dataset (stratified by acquisition day), the model achieved $89.71\% \pm 5.51\%$ overall pixel accuracy and $80.96\% \pm 13.37\%$ mean tissue Dice coefficient (see Tab.~\ref{tab:test_set_comprehensive_performance_final}). 
%
%
\begin{table}
\centering
\caption{Test dataset performance (mean $\pm$ std, in \%).}
\label{tab:test_set_comprehensive_performance_final}
\small
\renewcommand{\arraystretch}{1.3}
\begin{tabular}{lcc}
\hline
\textbf{Metric} & \textbf{Dice (\%)} & \textbf{IoU (\%)} \\
\hline
Pixel Accuracy & - & $89.7 \pm 5.5$ \\
Mean tissue DSC & $81.0 \pm 13.4$ & - \\
\hline
Background & $92.4 \pm 9.4$   & $87.1 \pm 14.2$ \\
Tissue     & $88.6 \pm 4.4$   & $79.9 \pm 7.1$  \\
OS         & $84.9 \pm 9.7$   & $74.8 \pm 13.1$ \\
Vaginal    & $69.4 \pm 32.6$  & $60.7 \pm 30.5$ \\
\hline
\end{tabular}
\end{table}
%
%

\begin{figure}[h!] 
\centering
\includegraphics[width=0.7\linewidth]{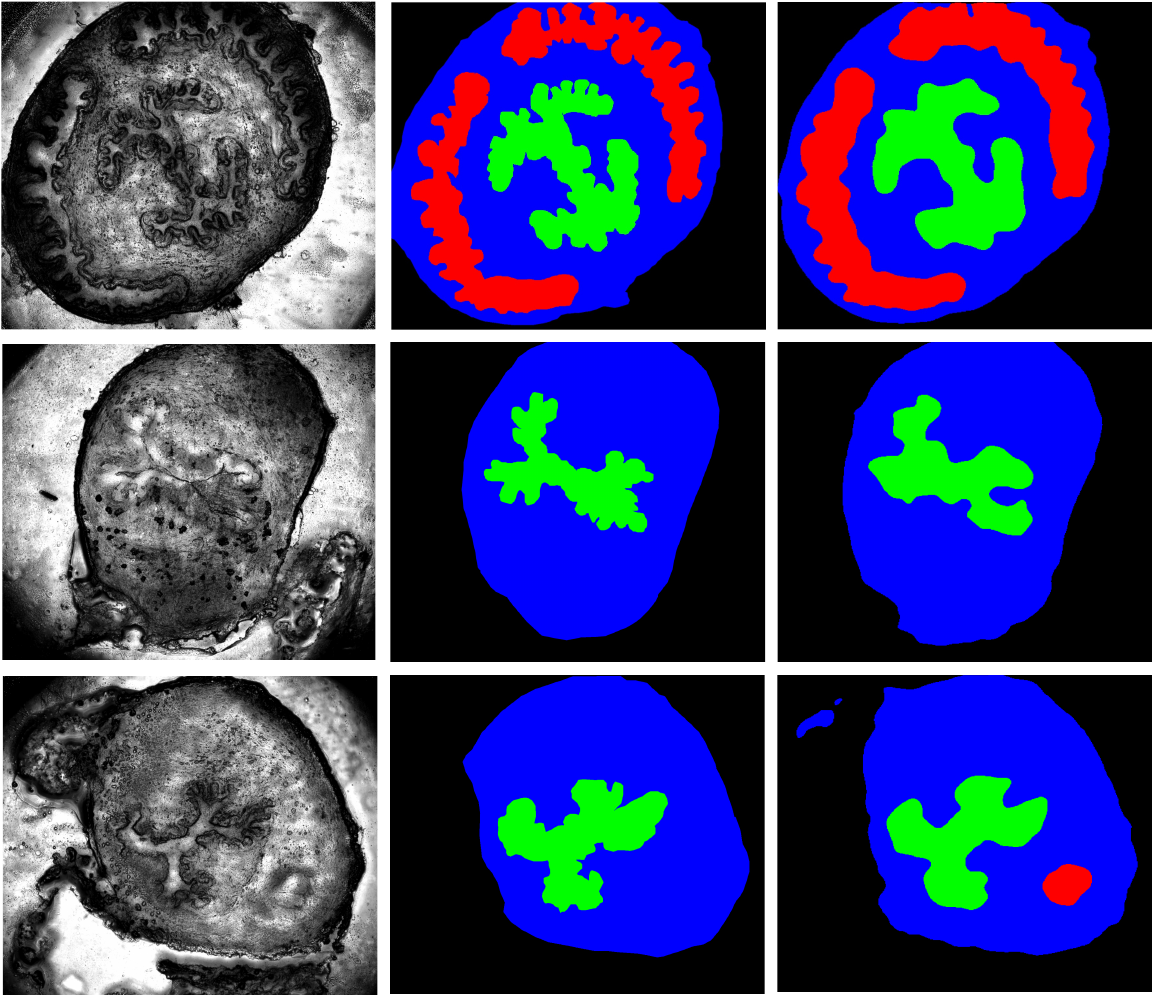}
\caption{Successful segmentation cases. 
Columns - \textbf{Left}: 
total 
intensity (M$_{11}$ images); \textbf{Middle}: 
ground truth images (blue=tissue, green=OS, red=vaginal wall); \textbf{Right}: 
model predictions \textbf{(top)} - clear boundaries; \textbf{(middle)} - correct absence of vaginal walls; \textbf{(bottom)} - detection of structure missed in manual annotation.}
\label{fig:m11_results_good}
\end{figure}

Per-class performance showed accurate segmentation for background (DSC: $92.37\% \pm 9.42\%$), tissue ($88.63\% \pm 4.44\%$), and internal os ($84.85\% \pm 9.71\%$). The vaginal wall segmentation exhibited higher variance (DSC: $69.41\% \pm 32.61\%$) due to its anatomically variable presence across sectioning depths and preparation artifacts, as seen in Figure~\ref{fig:m11_filter}. The overlap metric IoU (
Intersection over Union)   demonstrates similar trends.
The representative model predictions are shown in Figure~\ref{fig:m11_results_good}. The model accurately segments well-preserved samples (top), correctly predicts vaginal wall absence (middle), and can detect low-contrast structures missed during manual annotation (bottom), suggesting learned sensitivity to subtle polarimetric features. 

\begin{figure}[h!]
\centering
\includegraphics[width=0.7\linewidth]{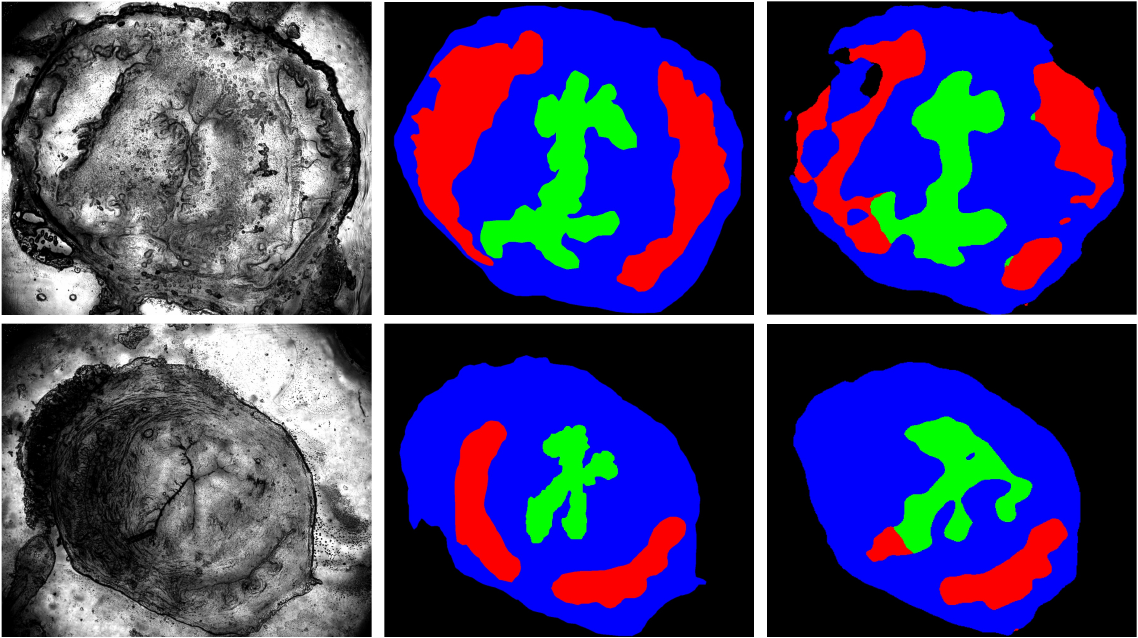}
\caption{Failure cases. Columns - \textbf{Left}: total 
intensity (M$_{11}$ images); \textbf{Middle}:
ground truth images 
; \textbf{Right}: model predictions \textbf{(top)} - segmentation errors because of tissue damage during preparation;  \textbf{(bottom)}  - 
under-performance of tissue segmentation because of
poor illumination.}
\label{fig:m11_results_bad}
\end{figure}

Fig.~\ref{fig:m11_results_bad} illustrates the failure cases: the segmentation errors in damaged tissue (top) reflect ground truth ambiguity, while 
segmentation in poorly illuminated regions (bottom) shows degraded performance under low signal-to-noise conditions.

\section{Discussion}

We demonstrated automated tissue segmentation from Mueller 
microscopy measurements using only the intensity image (M$_{11}$ element) as an input data. By leveraging transfer learning from ImageNet-pretrained ResNet-34 weights, the approach achieved robust segmentation ($89.71\%$ pixel accuracy, $80.96\%$ mean tissue DSC) with only 70 total samples (49 training, 11 validation, 10 test). This small dataset requirement addresses a critical bottleneck in biomedical applications of imaging polarimetry, where acquiring and annotating specialized tissue samples is labor-intensive and often limited by availability of specimens.

The success of our approach with limited training dataset demonstrates that pretrained encoders effectively transfer natural image features to polarimetric intensity patterns, requiring minimal domain-specific fine-tuning. This contrasts with training from scratch, which typically demands hundreds to thousands of annotated samples~\cite{ronneberger2015u}. The single-channel intensity-based approach further reduces preprocessing complexity compared to full MM decomposition methods~\cite{chae2025machine}, enabling rapid deployment for biomedical workflows where annotation resources are scarce. Importantly, because our method relies only on the M$_{11}$ total intensity images, it is extensible to other tissue types using other imaging modalities.

The model's ability to detect low-contrast vaginal tissue missed during manual annotation (Fig.~\ref{fig:m11_results_good}, bottom) suggests learned sensitivity to polarimetric intensity patterns beyond subjective human interpretation, reducing inter-observer variability. However, the high variance in vaginal wall segmentation ($69.41\% \pm 32.61\%$ DSC) reflects genuine biological variability—the structure's presence depends on sectioning plane and depth, introducing fundamental uncertainty even in ground truth labels.
The primary limitation of our small-sample approach is sensitivity to domain shift: samples with significantly different morphology may require additional training data for adaptation. Failure cases remain tied to sample preparation quality and imaging conditions (Fig.~\ref{fig:m11_results_bad}).

The framework's accessibility—requiring only ~50 annotated samples, normalized intensity images, and standard U-Net architecture enables straightforward extension to other tissue types and imaging modalities. Our graphical annotation tool and trained models are publicly available
, providing a practical pipeline for the analysis of MM images of tissues in resource-limited settings where large-scale annotation is impractical.


\section*{Acknowledgments}
SC, TN acknowledge support from the EUR BERTIP (ANR 18EURE0002, Program France 2030) and European Cooperation in Science and Technology (COST) actions CA21159 PhoBioS and CA23125 TETRA. JRR acknowledges support from the National Science Foundation (NSF) Award $\#$DMR-1548924. JRR, AA, DG, JZP acknowledge support from the NSF Award $\#$16484510.

\section*{Disclosures}
The authors declare no conflicts of interest.

\section*{Data Availability}
Data  are not publicly available at this time but may be obtained from the authors upon
reasonable request.

\section*{Code Availability}
\label{sec:m11_filter_code_availability}
The 
GUI for mask selection, the code used to train the model and the trained model (Python and MATLAB implementations) are available at: 
\url{https://github.com/chaetries/mmTissueFilter}.

\bibliographystyle{unsrt} 
\bibliography{report}

\end{document}